\documentclass[preprint,12pt]{elsarticle}
\pdfoutput=1 

\usepackage{graphicx}
\usepackage{amssymb}

\usepackage{cleveref}
\usepackage{lmodern}
\usepackage{rotating}
\newcommand{\ba}{\begin{eqnarray}}
\newcommand{\ea}{\end{eqnarray}}
\newcommand{\be}{\begin{equation}}
\newcommand{\ee}{\end{equation}}
\newcommand{\benn}{\begin{equation*}}
\newcommand{\eenn}{\end{equation*}}
\newcommand{\eps}{\epsilon}

\def\du#1{\underline{\underline{#1}}}
\def\underbrace#1{\underbrace{\underbrace{#1}}}
\usepackage{cuted}
\usepackage{bigdelim}

\usepackage[usenames, dvipsnames]{color}
\usepackage{amsmath}

\usepackage{graphicx}
\usepackage{tensor}
\usepackage{wrapfig}
\usepackage{float}
\usepackage[normalem]{ulem} 

\usepackage{multirow}

\usepackage{subfig}

\usepackage{mathrsfs}
\usepackage{mathtools}
\usepackage{amsmath}
\usepackage{amsthm}
\usepackage[makeroom]{cancel}

\allowdisplaybreaks
\usepackage{cases}
\usepackage{natbib}
\definecolor{LightBlue}{RGB}{28,117,188}
\definecolor{DarkBlue}{RGB}{5,64,112}

\definecolor{LightGreen}{RGB}{86,154,115}
\definecolor{DarkGreen}{RGB}{70,121,92}

\usepackage{rotating}
\usepackage{tabularx}
\usepackage[labelfont=bf]{caption} 
\usepackage{ragged2e}

\usepackage{array}
\usepackage{makecell}

\usepackage{subfiles}

\graphicspath{{figures/}{../figures/}}

\numberwithin{equation}{section}

\journal{-}

\begin{document}

\author{D. Eeltink}
\author{A. Armaroli}
\author{M. Brunetti}
\author{J. Kasparian}
\cortext[mycorrespondingauthor]{Corresponding author}
\ead{jerome.kasparian@unige.ch}

\address{Group of Applied Physics and Institute for Environmental Sciences, University of Geneva, blvd. Carl-Vogt 66, Geneva, Switzerland}

\begin{frontmatter}


\title{Reconciling different formulations of viscous water waves\\ and their mass conservation}



\begin{abstract}

The viscosity of water induces a vorticity near the free surface boundary. The resulting rotational component of the fluid velocity vector greatly complicates the water wave system. Several approaches to close this system have been proposed. Our analysis compares three common sets of model equations. The first set has a rotational kinematic boundary condition at the surface. In the second set, a gauge choice for the velocity vector is made that cancels the rotational contribution in the kinematic boundary condition, at the cost of rotational velocity in the bulk and a rotational pressure. The third set circumvents the problem by introducing two domains: the irrotational bulk and the vortical boundary layer. This comparison puts forward the link between rotational pressure on the surface and vorticity in the boundary layer, addresses the existence of nonlinear vorticity terms, and shows where approximations have been used in the models. Furthermore, we examine the conservation of mass for the three systems, and how this can be compared to the irrotational case.
\end{abstract}

\begin{keyword}
Vorticity \sep Viscosity \sep Gravity surface waves \sep Mass conservation


\end{keyword}

\end{frontmatter}



\section{Introduction}\label{sec_Intro}
 
For many hydrodynamic problems, the Navier-Stokes (NS) equation can be simplified by considering the fluid as inviscid and incompressible. Within these approximations, the inviscid water wave problem reduces to solving the Laplace equation within the bulk, together with the boundary conditions for the free surface and the rigid bottom. This set of equations for the surface elevation $\eta$ and the velocity potential $\phi$ can be used as the starting point to obtain Nonlinear Schr\"odinger (NLS) equation-like propagation models, or can serve as a basis for higher order spectral methods \citep{West1987,Dommermuth1987,Ducrozet2016}.

While this approach is sufficient in many situations, in reality, water is a viscous medium. The molecular viscosity accounts for wave damping but also plays a more intricate role  for instance in downshifting of the spectrum \cite{Carter2016}, in stabilizing the Benjamin-Feir instability \citep{Segur2005}, or can serve as a model for the eddy viscosity in the case of breaking waves \citep{Tian2010,Longuet-Higgins1992}. In domains such as the dissipation of swells \citep{Babanin2012},  visco-elastic waves propagating in ice \citep{Bennetts2012}, or the motion of very viscous fluids such as oil spills, considering viscosity is important. 

The inclusion of viscosity in the water wave problem has been proposed in several ways. One option, hereafter denoted as System A, used in \citet{Ruvinsky1991,Dias2008} (hereafter RFF and DDZ, respectively), includes the rotational part of the velocity vector into the kinematic boundary condition (KBC) at the surface. As the vorticity is assumed to be confined to only a thin boundary layer below the surface, the rotational velocity is zero in the bulk.

The second option, denoted System B  (\citet{Dommermuth1993}), makes a gauge choice for the velocity vector such that the rotational contribution in the  KBC disappears. The cost is however, that the rotational part of the velocity is nonzero in the bulk. Moreover, the pressure is split into rotational and irrotational parts too, introducing an additional equation for the rotational pressure that couples to the Navier-Stokes equation. This system is however fully nonlinear and makes no boundary layer approximations. 

The third option, denoted System C (\citet{Longuet-Higgins1969,Longuet-Higgins1992}), circumvents the difficulty of the rotational kinematic boundary condition by explicitly splitting the problem into two domains: the irrotational bulk and the vortical boundary layer. The former is shown to receive an additional pressure due to the weight of the latter.

The goal of this paper is to contrast and compare these models, offering a physical understanding of their mathematical differences. In this comparison (Sections 2 and 3), we map out the relation between the rotational part of the velocity vector and the pressure. In addition, we highlight where linearization is applied in order to close the systems and we explore the importance of nonlinear vortical terms. We point out in which situation a given model is more appropriate than another.  Furthermore, we examine the conservation of mass conditions for each system and show how this should be interpreted with respect to the irrotational system, further highlighting the relation between vorticity and pressure (Section 4). Finally, we summarize our findings (Section 5).

\begin{figure}
\centering
\includegraphics[width=0.70\textwidth]{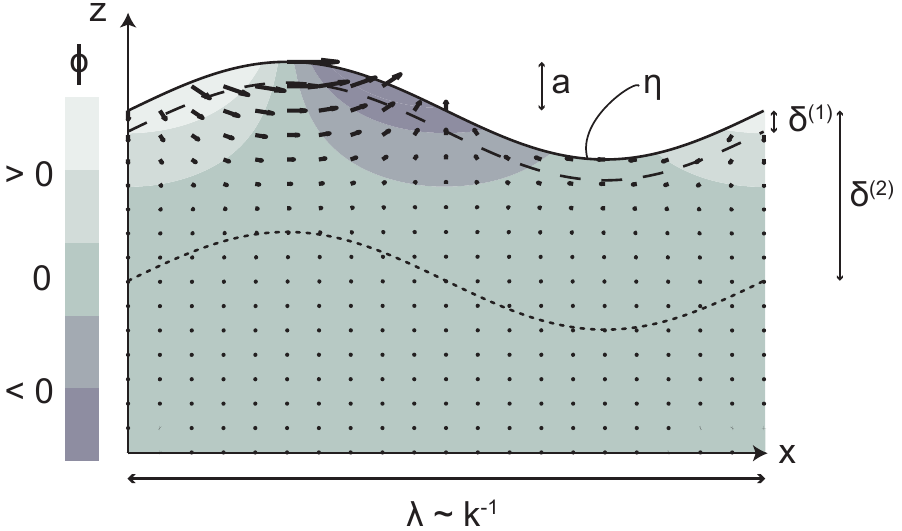}
\caption{The water wave problem and its relevant length scales. The velocity potential $\phi$ is indicated by the color-scale. The arrows indicate the velocity vector $\vec{u}$, following the gradient of $\phi$. The surface elevation $\eta (x,z)$ is indicated by the black line. The vortical boundary layer $\delta$ is displayed for two different regimes, indicated by the superscript: (1) $\delta < a$, or   $\frac{a}{\delta}=\frac{\eps}{\delta k}>1$ (dashed line) and (2) $\delta > a$ or $\frac{a}{\delta}=\frac{\eps}{\delta k}<1$ (dotted line).}\label{fig_LengthScales}
\end{figure}
\section{Boundary Conditions}
The physics of the water wave problem is defined by the boundary conditions. Figure \ref{fig_LengthScales} depicts the 2D water wave problem and its relevant quantities and length scales. The surface elevation $\eta(x,t)$ is denoted by the black line, the arrows are the local velocity vectors and the color scale refers to the value of the velocity potential $\phi(x,z,t)$, based on a linear wave \cite{Lamb1932}. 
\subsection{Shear stress: viscosity and vorticity} \label{sec:ShearStress}

For non-viscous waves the velocity vector $ \vec{u}$ is irrotational and can be written as the gradient of a potential field: 
\be \label{eqn_pot}
\vec{u} = \nabla \phi.
\ee
However, in the presence of viscosity, the continuity of tangential stresses at the free surface can only be fulfilled by rotational motion of the fluid \cite{Longuet-Higgins1992a, Lundgren1999}. 

The stress tensor in 2D tangential and normal components can be written as:
\begin{gather}
 \du{\sigma} = \begin{bmatrix} 2\mu \vec{u}^s_s - P & \mu(\vec{u}^n_s+\vec{u}^s_n)\\ \mu (\vec{u}^s_n+\vec{u}^n_s) & 2\mu \vec{u}^n_n-P\end{bmatrix} ,
\end{gather}
\noindent where $\mu$ is the dynamic viscosity, and $P$ the pressure.  We consider gravity waves in our analysis and therefore ignore the effect of the surface tension, as it is negligible. Note the different meanings of $s$ and $n$ as subscript or superscript. Here and in the following, subscripts denote the partial derivatives and superscripts denote the components of a vector, where $s$ denotes tangential and $n$ normal. The tangential stress component is

\begin{align}
       \sigma^s &= \tau^s = \mu (\vec{u}^s_n+\vec{u}^n_s) \label{eqn_sigmas},
\end{align}
\noindent where $\tau$ is the deviatoric stress tensor.

\begin{figure}[ht]
    \centering
\includegraphics[width=0.35\textwidth]{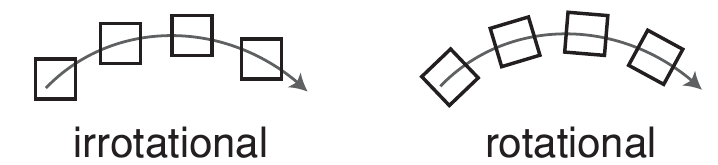}
\caption{In an irrotational flow there can be circular paths for the fluid, but each individual fluid particle does not rotate.} \label{fig_rot}
\end{figure}

Since $\mu$ in air is much smaller than in water, the shear stress must vanish at the surface, implying $\vec{u}^s_n=-\vec{u}^n_s$. There is thus no relative distortion to the fluid-particle, which due to the curvature of the interface, results in a rotational flow \cite{Longuet-Higgins1992a}. See Fig.~\ref{fig_rot} for an illustration of a rotational and an irrotational flow.

For the free surface water wave problem, viscosity therefore directly implies vorticity, \textit{i.e.} a rotational flow. That is, it is unphysical to have a viscid, irrotational flow \citep{Lundgren1999}.

For a rotational flow, the Helmholtz decomposition is used to split the velocity field into an irrotational part $\nabla \phi$, and a rotational, solenoidal ($\nabla \cdot \vec{U}=0$) part, $\vec{U}$:

\be \label{eqn_HH}
\vec{u}= \nabla \phi + \vec{U}= \nabla \phi + \nabla \times \vec{A}.
\ee

Since $\vec{u}$ only contains components in the $x,z$ plane, $\vec{A}$ only has a component in the $y$ direction, and can therefore be treated as a scalar, $A$. The velocity vector $\vec{u}  = (u,w)$ can explicitly be written as
\begin{align} \label{eqn_def_vPhiA}
u &= \phi_x + U = \phi_x - A_z, &  \vec{u}^s &= \phi_s +  \vec{U}^s \\
w &= \phi_z +  W = \phi_z +  A_x , & \vec{u}^n &= \phi_n +  \vec{U}^n.
\end{align}

Using the Helmholtz decomposition requires finding a harmonic function $\phi$ that satisfies $\nabla^2\phi=0$ and a solenoidal field $\vec{U}$ that satisfies the NS equations \citep{Joseph2006}. Therefore, certain transfers of irrotational flow from the $\nabla\phi$ term to the $\vec{U}$ are allowed, keeping Eq.~(\ref{eqn_HH}) valid. That is, while $\nabla \phi$ is irrotational (since the curl of a gradient is always 0), $\vec{U}$ can include an irrotational part on top of its rotational part \citep{Dommermuth1993}. 

This implies that one cannot assume a-priori that $\nabla\phi$ contains the full irrotational velocity potential: its value differs for different gauge choices.

\begin{figure*}[ht]
\centering
\includegraphics[width=0.99\textwidth]{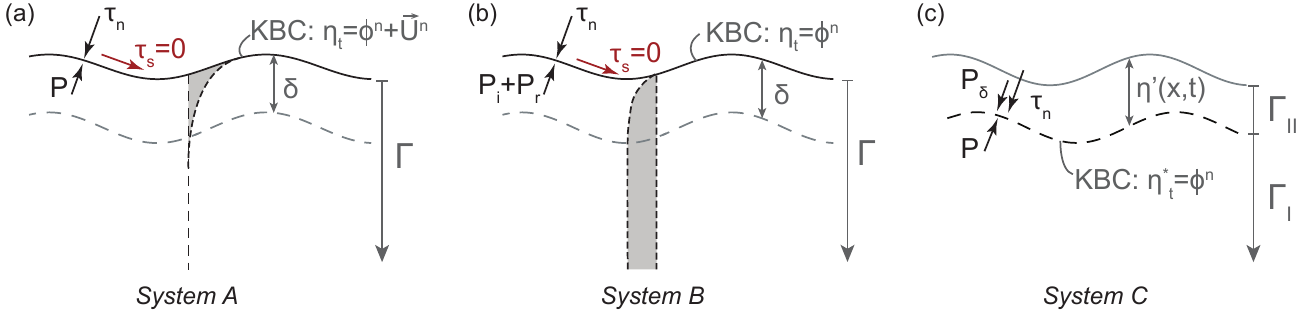}
\caption{(a) System A: The boundary conditions include the balance of normal stress, the vanishing of the shear stress, and the rotational KBC. The grey shaded area indicates that $\vec{U}^n$ decays over a characteristic length $\delta$. (b) System B: The boundary conditions include the splitting of the pressure for the normal stress, the vanishing of the shear stress, and an irrotational KBC. The grey shaded area indicates that while $\vec{U}^n=0$ at the surface, this is not the case within the bulk. (c) System C: The domain is split into two sub domains separated by $\eta^*$: the bulk ($\Gamma_1$)  and the vortical boundary layer ($\Gamma_2$). The boundary conditions for the irrotational domain $\Gamma_1$ on $\eta^*$ are the continuity of normal stress, with an added pressure due to the weight of $\Gamma_{II}$, and an irrotational KBC. The shear stress does not need to vanish at $\eta^*$.}\label{fig_3situations}
\end{figure*}

\subsection{Kinematic boundary condition} \label{sec_KBC}
 The KBC ensures that fluid particles on the free surface always remain there. As it is a key ingredient for the rest of our discussion,  we shall derive it explicitly. We can describe the surface elevation by $z = \eta(x,t)$, and let the level set $f(x,y,t)=0$ define the interface between air and water
\be \label{eqn_KBC_f}
f(x,z,t) \equiv  z - \eta(x,t).
\ee
\noindent Because $f=0$ on the interface at all times $t$, its material derivative, $D/Dt$, must be null
\be \label{eqn_KBC_matD}
\begin{split}
\frac{Df}{Dt} &\equiv \frac{\partial f}{\partial t} + \vec{u} \cdot\nabla f = 0 \qquad \textrm{on} \quad f = 0 .
\end{split}
\ee
\noindent To write the KBC in terms of $\eta$ and the velocity vector $\vec{u}=(u,w)$, inserting Eq.~(\ref{eqn_KBC_f}) into Eq.~(\ref{eqn_KBC_matD}) yields 
\be \label{eqn_KBC_u}
\frac{\partial \eta}{\partial t} = - u\frac{\partial \eta}{\partial x} + w  \qquad \textrm{at} \quad z = \eta  .
\ee
\noindent in Cartesian coordinates. In normal and tangential coordinates  Eq.~(\ref{eqn_KBC_matD}) reads:
\be \label{eqn_KBC_u2}
\frac{\partial \eta (x,t)}{\partial t}  = \vec{u} \cdot \hat{n} |\nabla f|  = \vec{u}^n  \sqrt{1+\eta_x^2} ,
\ee
\noindent where the unit normal vector is defined as $\hat{n}= \nabla f / |\nabla f|=(-\eta_x,1)/\sqrt{\eta_x^2+1}$, and points outwards. Using $\eta(s,t)$, the KBC can be written as:
\be \label{eqn_KBC_u3}
\frac{\partial \eta (s,t)}{\partial t}  =  \vec{u}^n .
\ee

\noindent The latter expression corresponds to the intuitive image that the deformation of the surface, \textit{i.e.} the change of $\eta$ in time, is equal to the normal component of the velocity vector $\vec{u}^n$ pushing the surface either inwards or outwards.

\subsection{Dynamic boundary condition}\label{sec_DBC}
In addition to the continuity of shear stress, the normal stress must also be continuous over the boundary between water (w) and air (a) $\sigma^{n,\textrm{w}}=\sigma^{n,\textrm{a}}$, yielding the dynamic boundary condition (DBC). We can write
\be
 \sigma^n = \tau^n - p^n = 2\mu \vec{u}^n_n-P \label{eqn_sigman}  \qquad \textrm{at} \quad z = \eta  , \\
\ee

\noindent where $P$ is the pressure. 

\subsubsection{Irrotational Flow}
For an irrotational flow, the continuity of normal stress reduces to the continuity of pressure: $P^{\textrm{w}}=P^{\textrm{a}}$. For an irrotational, incompressible flow, integrating the NS equation in space leads to the Bernoulli equation:
\be \label{eq_Bernoulli1}
P^{\textrm{w}} = -\rho\left(\phi_t + \frac{1}{2} (\nabla \phi)^2 +  g\eta\right) + c(t) = P^{\textrm{a}}  \qquad \textrm{at} \quad z = \eta  .
\ee
Choosing the arbitrary Bernoulli function $c(t) = P^{\textrm{a}} $ gives the DBC:
\be \label{eq_Bernoulli}
\phi_t + \frac{1}{2} (\nabla \phi)^2 +  g\eta = 0  \qquad \textrm{at} \quad z = \eta  .
\ee

\subsubsection{Rotational Flow}

Due to the Helmholtz decomposition, the situation is more complicated for a rotational fluid. Again we start from the continuity of normal stresses:
\be  
2 \mu \vec{u}^n_n - P^{\textrm{w}} = -P^{\textrm{a}}  \qquad \textrm{at} \quad z = \eta  .
 \ee
Using the Helmholtz decomposition gives
\be  \label{eqn_press}
2 \nu \left(\phi_{nn} + \vec{U}^n_n\right) -P^{\textrm{w}} = -P^{\textrm{a}}  , 
\ee
\noindent where  $\nu=\mu/\rho$ is the  kinematic viscosity. The pressure $P^{\textrm{w}}$ can be obtained from the Navier-Stokes equation for an incompressible, Newtonian fluid :
\be \label{eqn_NS_incompr}
\vec{u}_t + \vec{u} \cdot \nabla \vec{u}  = - \frac{1}{\rho}\nabla P^{\textrm{w}} + \vec{g}  + \nu \nabla^2 \vec{u} .
\ee

\noindent Using the relation  $\vec{u} \cdot \nabla \vec{u} =\frac{1}{2} (\nabla \vec{u})^2 + \omega \times \vec{u}  $ in  Eq.~(\ref{eqn_NS_incompr}), and again integrating in space, yields
\begin{multline} \label{eqn_DBC_P}
\underbrace{ \phi_t + \frac{1}{2} (\nabla \phi)^2 +  \frac{P^{\textrm{w}}}{\rho}  + g\eta+c(t)}_\textrm{Bernoulli} = \\\underbrace{ - \int_{\eta-\delta}^{\eta} \left( \vec{U}_t - \nu \nabla^2 \vec{U}     - \omega \times \vec{U} \right) dz   - \frac{1}{2}\vec{U}^2}_\textrm{Vortical layer} \\
\underbrace{+\phi_n \vec{U}^n + \phi_s \vec{U}^s}_\textrm{Mixed NL terms} .\qquad 
\end{multline}

\noindent Inserting Eq.~(\ref{eqn_DBC_P}) into the balance of normal stress (Eq.~(\ref{eqn_press})) gives the DBC. Here and in the following, we always choose the arbitrary integration function $c(t)$ such that it cancels $P^{\textrm{a}}$. Therefore, we omit the $w$ index for $P^{\textrm{w}}$ from now on. Compared to the inviscid water wave problem, the rotational part of the velocity $\vec{U}$ is a third unknown in addition to  $\phi$ and $\eta$. To close the system different routes can be taken. We shall now review three of these.
\linebreak

\section{Three complementary views}  \label{sec_ThreeSystems}
The three systems introduced in the Introduction, are summarized in Figure \ref{fig_3situations} and Table \ref{tab:summary}.

In System A (Figure \ref{fig_3situations}A), the velocity has a rotational part (Eq.~(\ref{eqn_HH})), and therefore a rotational KBC. The vortical component of the velocity, indicated by the grey shaded area, is nonzero at the boundary and rapidly decays to zero, over a typical length $\delta$: the width of the vortical boundary layer. There are different ways to close this system. One, described in RFF \citep{Ruvinsky1991}, is to impose a separate boundary condition for the rotational part of the velocity vector $\vec{U}$, based on the vorticity equation. The second, introduced in DDZ \citep{Dias2008}, is to write $\vec{U}$ in potential flow terms, using expressions obtained from the linearized equations. 

In System B (Figure \ref{fig_3situations}B), as presented in \citet{Dommermuth1993}, an additional boundary condition $\vec{U}^n = 0$ is imposed. This is in effect a gauge choice in the Helmholtz decomposition, and leads to an irrotational velocity at the surface, Eq.~(\ref{eqn_pot}). The price for the irrotational KBC is that the normal stress continuity boundary condition now pertains to the sum of the irrotational and the rotational parts of the pressure: $P=P_{\text{i}}+P_{\text{r}}$. Moreover, the vortical component of the velocity (the grey shaded area) is nonzero in the bulk of the fluid.  

In System C  (Figure \ref{fig_3situations}C), derived by \citet{Longuet-Higgins1969,Longuet-Higgins1992}, the problem is split into two domains: an irrotational bulk ($\Gamma_I$), and a rotational boundary layer ($\Gamma_{II}$). The equations are solved for $\Gamma_I$. Since there is no interface with air for $\Gamma_I$, the shear stress on the boundary does not have to vanish, and the viscous fluid in this domain can therefore be irrotational. The weight of $\Gamma_{II}$ induces a pressure $P_\delta$ on the top boundary of $\Gamma_I$. An expression for $P_\delta$ is derived in terms of the mass-flux of $\Gamma_{II}$, caused by the rotational part of the velocity ($\vec{U}$).

\subsection{Shared equations}
For all three systems, the continuity equation in the bulk $\nabla \cdot \vec{u} =0$ yields the Laplace equation for the velocity potential:

\begin{equation}  \label{eqn_cont}
   \nabla^2~\phi~=0 . 
\end{equation}

They also share the same bottom boundary condition in the deep water limit: $\phi_z~\rightarrow ~0 $ as $ z~\rightarrow  ~-\infty $. 

We shall now discuss each system in detail. We shall first provide the remaining equations for each system, and then discuss each equation. All results are also summarized in Table \ref{tab:summary}.

\subsection{System A} \label{sec_sys_Ruv}
The boundary conditions at the free surface boundary, $z=\eta$  can be written as

\begin{subequations}\label{sys_A_total}
\begin{align}
&\eta_t + \eta_x\phi_x + \underbrace{\eta_x U}_{\text{NLV1}} =  \phi_z + W & \text{KBC} \label{sys_A_KBC}  \\ 
&\sigma^n \equiv 2\mu \phi_{zz} - P= -P^{\textrm{a}} & \perp-\text{stress} \label{sys_A_Norm} \\
& P=- \rho\left( \phi_t  +\frac{1}{2}(\nabla\phi)^2 + g\eta + \underbrace{\phi_xU}_{\text{NLV2}}\right) + c(t) & \text{NS} \label{sys_A_NS}\\ 
& \rightarrow \quad  \phi_t  +\frac{1}{2}(\nabla\phi)^2 + g\eta = - 2\nu\phi_{zz} -\underbrace{\phi_xU}_{\text{NLV2}}  & \text{DBC} \label{sys_A_DBC} 
\end{align}
\end{subequations}

\noindent where NS refers to the space-integrated Navier-Stokes equation evaluated at the surface. The arrow indicates that inserting the latter into the balance of normal stress Eq.~(\ref{sys_A_Norm}) yields the DBC. Recall that we use $c(t)$ to cancel $P^{\textrm{a}}$.
\begin{itemize}
    \item In the KBC, the fluid is considered rotational (see Fig.~\ref{fig_3situations}A). Therefore the Helmholtz notation is used, Eq.~(\ref{eqn_KBC_u}),  giving Eq.~(\ref{sys_A_KBC})\footnote{
The vortical part decreases rapidly over a distance $\delta=\sqrt{\frac{2\nu}{\omega}}$, as demonstrated in \citet{Lamb1932} section 348, where the stream function $\Psi$ corresponds to $A$ in Eq.~(\ref{eqn_def_vPhiA}).}.  Like Eq.~(\ref{eqn_KBC_u3}), this can also be written as
\be \label{eqn_KBC_A2}
\eta(s,t)_t = \phi_n + \vec{U}^n. 
\ee
\item The normal stress boundary condition (Eq.~(\ref{sys_A_Norm})) is Eq.~(\ref{eqn_press}), neglecting the term $\vec{U}^n_n$,  because $\vec{U}^n/\phi_{n} \sim k\delta \ll 1$. In addition, the approximation $\partial / \partial n \approx \partial / \partial z$ is used.

\item To obtain the pressure, in Eq.~(\ref{eqn_DBC_P}), we retain terms of $\mathcal{O}(\eps\delta k)$, that is only the term $\phi_s \vec{U}^s \approx \phi_x U$, yielding:

\be \label{A_eqn_pressure}
-\frac{P}{\rho}  = \underbrace{\phi_t + \frac{1}{2}(\nabla \phi)^2}_\textrm{Dynamic pressure}  \underbrace{+g\eta}_\textrm{Static pressure} + \underbrace{\phi_x U}_\textrm{NL vort. term}.
\ee

\item The DBC (Eq.~(\ref{sys_A_DBC})) is obtained by inserting the pressure from the NS equation  (Eq.~(\ref{sys_A_NS})) into the balance of normal stress (Eq.~(\ref{sys_A_Norm})).

\end{itemize}
We discuss three variations of System A:

\subsubsection{Ruvinsky, Feldstein and Freidman, 1991}
The first method to close the system is developed by RFF (\citep{Ruvinsky1991}), who insert the Helmholtz decomposition in Eq.~(\ref{eqn_NS_incompr})  and rewrite the vector products using $\nabla \phi \cdot \nabla (\nabla \phi) = \nabla \frac{1}{2}(\nabla \phi)^2$, giving
\begin{multline}
\nabla\phi_t + \vec{U}_t + \nabla \frac{1}{2}(\nabla \phi)^2+ ((\nabla \phi + \vec{U}) \cdot \nabla) \vec{U} + (\vec{U}\cdot \nabla) \nabla \phi   = \\
- \frac{1}{\rho}\nabla P + \vec{g}   + \nu \nabla^2 \vec{U} .
\end{multline}
Subsequently, they move the gradient terms to the LHS, and on the RHS integrate along the vertical direction, in a small range near the surface ($\eta-\delta,\eta$). Taking the normal component, and assuming that $\vec{U}=0$ outside the boundary layer, $z \in (\eta-\delta,\eta)$ they obtain
\begin{multline} \label{eqn_Ptot_Ruv}
\phi_t + \frac{1}{2}(\nabla \phi)^2 + g\eta   + \frac{P}{\rho}  + c(t)   = \\
- \underbrace{\int_{\eta-\delta}^{\eta}\hat{n}\cdot \left( \vec{U}_t +  ((\nabla \phi + \vec{U}) \cdot \nabla) \vec{U} + (\vec{U}\cdot \nabla) \nabla \phi - \nu \nabla^2 \vec{U} \right) dn}_\textrm{Vorticity terms in the boundary layer}.
\end{multline}
Performing the dimensional analysis they show that the integral terms on the RHS are of higher order in steepness $\eps = ak$ and $\frac{\delta}{a} \ll 1$, and are therefore neglected. Inserting into the balance of normal stress gives the DBC. Additionally, the vorticity equation ($\omega_t+(\vec{u}\cdot\nabla)\omega = \nu \nabla^2 \omega$ ) is evaluated at the boundary, yielding a vortical boundary condition (VBC), see \citep{Ruvinsky1991} for details. Together with the KBC, this gives the following system: 

\begin{subequations}\label{sys_RFF_total}
\begin{align}
&\eta_t + \eta_x\phi_x  =  \phi_z + W & \text{KBC} \label{sys_RFF_KBC}  \\ 
&\phi_t  +\frac{1}{2}(\nabla\phi)^2 + g\eta = - 2\nu\phi_{zz} & \text{DBC}\label{sys_RFF_DBC}  \\
&W_t=2\nu \phi_{zxx} & \text{VBC}\label{sys_RFF_VBC}
\end{align}
\end{subequations}

\noindent where we ignore the surface tension. 

\citet{Jarrad2001} perform a linear stability analysis on the RFF system \citep{Ruvinsky1991}, finding growing modes, while the only physical modes are decaying ones. However, when rewriting the VBC as in \citet{Tian2010}, only the DBC and KBC remain, and it is equivalent to the DDZ system (see next section). Performing a linear stability analysis on this system yields physical eigenvalues, with a $-2\nu k^2$ damping rate.

\subsubsection{Dias, Dyachenko and Zakharov, 2008}

The second method to close the system is described in DDZ (\citep{Dias2008}). The vortical velocity component $W$ is expressed in terms of $\phi$ or $\eta$. These expressions can be found using on the linear water wave problem. Subsequently, it can be conjectured that these expressions also hold in the nonlinear system, giving:  
\begin{subequations}\label{sys_DDZ_total}
\begin{align}
&\eta_t + \eta_x\phi_x  =  \phi_z + \nu \eta_{xx} & \text{KBC} \label{sys_DDZ_KBC}  \\ 
&\phi_t  +\frac{1}{2}(\nabla\phi)^2 + g\eta = - 2\nu\phi_{zz} & \text{DBC} \label{sys_DDZ_DBC} 
\end{align}
\end{subequations}

\subsubsection{Nonlinear vortical version System A}
The most general version of System A includes the nonlinear vorticity terms labeled VNL1 and VNL2 in Eqs.~(\ref{sys_A_total}) that are neglected by RFF and DDZ systems. The relevance of these terms will be discussed in detail in Section \ref{sec_Disc_NLV}.

\subsection{System B} \label{sec_sys_Domm}
The boundary equations at $z=\eta$ for System B (\citet{Dommermuth1993}) can be written as : 
\begin{subequations}\label{sys_B_total}
\begin{align}
&\eta_t + \eta_x\phi_x =  \phi_z   & \text{KBC1} \label{sys_B_KBC1}\\
&\vec{U}^n = 0 &  \text{KBC2} \label{sys_B_KBC2}\\
&\sigma^n\equiv2\mu \frac{1}{\eta_x^2 + 1} (\phi_{zz} + W_z)+ \text{NL} -(P_{\text{rot}} + P_{\text{irr}})=P^{\textrm{a}}& \perp \text{-stress} \label{sys_B_Norm}  \\
&P_{\text{irr}} = -\rho\left(\phi_t  + \frac{1}{2}(\nabla\phi)^2 + g\eta \right) + c(t) & \text{NS}, P_{\text{irr}} \label{sys_B_Pirr}\\
 &\rightarrow \quad \phi_t  +\frac{1}{2}(\nabla\phi)^2 + g\eta = - 2\nu\frac{1}{\eta_x^2 + 1} \left(\phi_{zz}+W_z\right)- \frac{P_\textrm{rot}}{\rho}   & \text{DBC}\label{sys_B_DBC} 
\end{align}
\end{subequations}
Where NL indicates nonlinear terms, which can be found in \citep{Dommermuth1993}. 

\begin{itemize}
    \item For the KBC, the additional boundary condition $\vec{U}^n=0$ is imposed at the surface (Eq.~(\ref{sys_B_KBC2})), rendering the KBC (Eq.~(\ref{sys_B_KBC1})) irrotational. This is achieved at the cost of $\vec{U}$  being non-zero in the bulk, as indicated in Figure \ref{fig_3situations}B. 
\item In order to obtain an additional equation to accommodate for the extra boundary condition, the pressure is also split in a rotational and irrotational part in the balance of normal stresses (Eq \ref{sys_B_Norm}):
\begin{align} 
    P &= P_{\text{irr}} + P_{\text{rot}}. \label{eqn_B_P}
\end{align}

\item The irrotational Bernoulli equation defines $P_{\text{irr}}$, the remaining part is denoted $P_{\text{rot}}$. 
\item  Like in System A, the DBC (Eq.( \ref{sys_B_DBC})) is obtained by inserting the equation for the pressure (Eq.~(\ref{sys_B_Pirr})) into the normal stress balance (Eq.( \ref{sys_B_Norm})). However, now, $P_{\text{rot}}$ remains unknown.

\item To obtain an equation for $P_{\text{rot}}$, the decomposition for the velocity (Eq.~(\ref{eqn_HH})), and for the pressure (Eq.~(\ref{eqn_B_P})) can be inserted into the viscous and rotational Navier-Stokes equation in the bulk, resulting in
\be\label{sys_B_Prot}  
\begin{split} 
\nabla  P_{\text{rot}} = U_t - ((\vec{U}+\nabla \phi)\cdot \nabla)\vec{U)}\\  
\qquad \qquad \qquad  -(\vec{U}\cdot \nabla\phi)\nabla\phi + \nu \nabla^2\vec{U}.& 
\end{split}
\ee
\noindent Taking the scalar product of the Navier-Stokes equation with the normal vector to the free surface, one can obtain a boundary condition for $\partial P_{\text{rot}} / \partial n$.  
\end{itemize}
Dommermuth's system retains all nonlinear terms without any approximations. This system was derived for studying the evolution of vortical cylinders moving from the bottom towards the water surface, where the vorticity is indeed not just limited to the boundary. The Helmholtz decomposition allows for a seamless transition between regimes with different Reynolds numbers.

\subsection{System C} \label{sec_sys_LH}
\citet{Longuet-Higgins1992,Longuet-Higgins1969} formally splits the problem into two domains: the irrotational bulk $\Gamma_{I}$, and the vortical boundary layer $\Gamma_{II}$, see Fig. \ref{fig_3situations}C. The equations for $\Gamma_I$ at its upper boundary $z=\eta^*$ are given by:
\begin{subequations}\label{sys_C_total}
\begin{align}
& \eta^*_t  + \eta^*_x\phi_x  =  \phi_z  & \text{KBC} \label{sys_C_KBC}\\
&\sigma_n \equiv 2\mu \phi_{zz} - P  = -\left(P_{\delta}+P^\text{a}\right) & \perp\text{-stress} \label{sys_C_Norm} \\
&P_{\delta}/\rho = g \eta' =  2\nu \phi_{zz} & \text{added } P_{\delta} \label{sys_C_Pdelta}\\
& P= -\rho \left(\phi_t  + \frac{1}{2}(\nabla\phi)^2 + g\eta^* \right)+c(t) & \text{NS} \label{sys_C_NS}\\
&\rightarrow \quad \phi_t  + \frac{1}{2}(\nabla\phi)^2+ g\eta^*  = -4\nu \phi_{zz} & \text{DBC} \label{sys_C_DBC}
\end{align}
\end{subequations}

\begin{itemize}
    \item In this configuration, the shear stress for the top boundary of $\Gamma_I$ , $\eta^*$,  does not have to vanish, as it does not have an interface with air. Therefore, the viscous fluid, and the KBC (Eq.~(\ref{sys_C_KBC})), can be treated as irrotational. This is in contrast to models like Refs. \citep{Joseph2006,Padrino2007}, which have the unphysical situation of an irrotational fluid with the free surface as top boundary: the vanishing of shear stresses at the free surface cannot physically occur in an irrotational fluid.
    \item The normal stress balance (Eq.~(\ref{sys_C_Norm})) receives an additional pressure $P_\delta$ due to the weight of the boundary layer above. 
    \item The Navier-Stokes equation (Eq.~(\ref{sys_C_NS})) is now evaluated at the  top of $\Gamma_I$: $\eta^*$.
    \item Again, combining the normal stress balance and the Navier-Stokes gives the DBC: Eq.~(\ref{sys_C_DBC}).
\end{itemize}

The domain of the boundary layer $\Gamma_{II}$ is considered not to have a constant thickness  $\delta$, but a variable height $\eta'(x,t)$. We can write for $\eta$:
\be \label{eqn_LH_etadef}
\eta =\eta^* + \eta'
\ee

Briefly, obtaining a function for $\eta'$ hinges on three critical observations:
\begin{enumerate}
    \item The boundary layer is \emph{defined} as a fluid region where there is a mass-flux, {\it i.e}. fluid flowing through the boundary, due to vorticity. This mass-flux can be written as
\be \label{eqn_LH_M}
M = \int_{\eta^*}^{\eta} \rho \vec{U}^s dn . 
\ee
\item The layer thickness is not constant in time, and a KBC can be written as

\be \label{eqn_LH_Un}
\frac{\partial \eta'}{\partial t}=\vec{U}^n = \int \vec{U}^n_n dn = - \int \vec{U}^s_s dn , 
\ee
\noindent where the last step is made using the fact that the divergence of $\vec{U}$ is null. Note that this corresponds to the change of the \textit{thickness} of the boundary layer in time. This is different from the motion of the upper boundary $\eta$, which would depend on the total normal velocity  $\vec{u}^n=\vec{U}^n+\phi_n$. 

Combining Eq.~(\ref{eqn_LH_M}) and Eq.~(\ref{eqn_LH_Un}) gives

\be \label{eqn_LH_deltat2}
\frac{\partial \eta'}{\partial t}=-\frac{1}{\rho}\frac{\partial M}{\partial s}\simeq-\frac{1}{\rho c}\frac{\partial M}{\partial t},
\ee

\noindent where in the \emph{linear} limit $\partial/\partial x \sim (1/c) \partial/\partial t$, with $c~=~\omega/k$  the phase speed, $\omega$ the orbital frequency and $k$ the wavenmber. This shows the intuitive relation that the difference between the mass-flux from one boundary at $s$ and the other at $s+ds$, $\frac{1}{\rho}\frac{\partial M}{\partial s}$, determines the fluid inflow into a slice, and must be equal to the change in height of the boundary layer, as displayed in Figure \ref{fig_MassCons}a. This point will be further discussed in Section \ref{sec_ConsMass}.

\item The total tangential stress $\tau^{s,\textrm{tot}}$ on the boundaries of the layer is equal to the mass transport
\be
 M_t = \tau^{s,\textrm{tot}}. 
\ee

\noindent Therefore, Eq.~(\ref{eqn_LH_deltat2}) can be written as
\be
\frac{\partial \eta'}{\partial t}=-\frac{\tau^{s,\textrm{tot}}}{\rho c} . 
\ee
\end{enumerate}

\noindent Assuming $\eta' \propto e^{i(kx - \omega t)}$, gives
\be
\eta'= -\frac{i \tau^{s,\textrm{tot}}}{\rho c \omega} , 
\ee
\noindent showing that $\eta'$ leads $\tau$ by 90$^{\circ} $. Since the shear stress at the surface must vanish, it is only the shear stress induced at the bottom of the vortical layer due to the viscous fluid motion of the irrotational bulk that contributes to $\tau^{s,\textrm{tot}}$: 
\be
\tau^{s,\textrm{tot}}= \tau^s= \mu (\vec{u}^s_n+\vec{u}^n_s)= 2\mu \vec{u}^n_s \approx  2 \mu \eta_{st} \approx 2 \mu \omega k \eta , 
\ee
\noindent where the last two steps are made assuming that $\eta$ is \textit{linear}. Now we can write
\be \label{eqn_LH_eta'}
\eta'(x,t) = - \frac{2i\mu k^2}{\rho \omega}\eta(x,t) =  2\nu \frac{k}{\omega^2} \phi_{zz}(x,t) . 
\ee
This layer produces an additional normal stress on its bottom boundary (denoted $\eta^*$), simply due to its own weight: $P_{\delta}=\rho g \eta' = 2\mu \phi_{zz}$.

Finally, this method does not rely on the size of $\delta$, or the relation between $\delta$ and $\eps$. Therefore, it is valid also for $\frac{a}{\delta}=\frac{\eps}{\delta k}\ll1$, unlike for instance the Stokes expansion \citep{Longuet-Higgins1953}. However, the expression for the pressure is made in the linear approximation, and consequently does imply $\eps \ll 1$

It is interesting to note that the KBC (Eq.~(\ref{sys_C_KBC})) and DBC (Eq.~(\ref{sys_C_DBC})) of System C are the same as those used by \citet{Wu2006}, namely:
\begin{subnumcases}{\label{sys_D_total}\hspace{-3em} z=\eta}
 \eta_t  + \eta_x\phi_x  =  \phi_z  & \hspace{-1em} KBC \label{sys_D_KBC}
\\
\phi_t  + \frac{1}{2}(\nabla\phi)^2+  g\eta   =   -  4 \nu \phi_{zz} & \hspace{-1em} DBC \label{sys_D_DBC}
\end{subnumcases}
This system is also suggested in the last sentence of the appendix in RFF as a simpler alternative for their system, without further explanation. Longuet-Higgins was able to give a physical underpinning for the irrotational KBC and the added factor 2 to the viscosity term $-2\nu \phi_{zz}$ in the DBC. However, while Eqs.~(\ref{sys_D_total}) refer to the surface elevation $\eta$ at the boundary between air and water, Eqs.~(\ref{sys_C_total}) refer to $\eta^*$, between the boundary layer and the bulk, as shown in Fig. \ref{fig_3situations}C. Nevertheless, since the absolute amplitude of the boundary is irrelevant in a deep water limit, the boundary condition follows the same motion as $\eta$, apart from the aforementioned phase-lag. Performing a multiple scales analysis on Eqs.~(\ref{sys_D_total}) gives the same viscous higher order Nonlinear Schr\"odinger equation (the Dysthe equation) for the propagation of the envelope as the DDZ system (Eq.~(\ref{eqn_propeqn_vort})).  

\begin{sidewaystable}
    \centering
     
    {\renewcommand{\arraystretch}{1.3}
    \scalebox{0.74}{\begin{tabular}{| c  c  c  c  c |}
    \hline
    &  & \textbf{System A} & \textbf{System B} & \textbf{System C}\\ 
         \cline{3-5}
$z< \textrm{upper boundary}$  &     &  {$\nabla^2\phi=0$} &  {$\nabla^2\phi=0$} &  {$\nabla^2\phi=0$} \\ 
 $ z = $lower boundary  &  &  {$\phi_z \rightarrow 0$}&  {$\phi_z \rightarrow 0$}&  {$\phi_z \rightarrow 0$} \\
\multirow{5}{*}{$z= \textrm{upper boundary}$ $\left\{\begin{tabular}{@{\ }l@{}}
    \\  \\  \\  \\  \\ 
  \end{tabular}\right.$}&  KBC &   $\eta_t + \eta_x\phi_x + \underbrace{\eta_x U}_{\text{NLV1}} =  \phi_z + W $ &  $\eta_t + \eta_x\phi_x =  \phi_z $ &  $ \eta^*_t  + \eta^*_x\phi_x  =  \phi_z $  \\
 &  KBC2 &  - &  $\vec{U}^n=0$ &  -  \\ 
&   $\perp$-stress   &  $\sigma^n \equiv 2\mu \phi_{zz} - P= -P^\text{a}$  &  $\sigma^n\equiv 2\mu \frac{1}{\eta_x^2 + 1} (\phi_{zz} + W_z)+\text{NL} -(P_{\text{rot}} + P_{\text{irr}})=-P^\text{a}$ &  $ \sigma^n \equiv 2\mu \phi_{zz} - P  = -(P_{\delta}+P^\text{a}) $\\ 
&   NS & $P= -\rho \left( \phi_t  +\frac{1}{2}(\nabla\phi)^2 + g\eta + \underbrace{\phi_xU}_{\text{NLV2}}\right) + c(t) $   &  $P_{\text{irr}}= -\rho\left(\phi_t  + \frac{1}{2}(\nabla\phi)^2 + g\eta\right)+ c(t)$ &  $P= -\rho \left(\phi_t  + \frac{1}{2}(\nabla\phi)^2 + g\eta^* \right)+ c(t)$ \\ 

&  $\rightarrow$ DBC &   $ \phi_t  +\frac{1}{2}(\nabla\phi)^2+ g\eta = - 2\nu\phi_{nn} -\underbrace{\phi_xU}_{\text{NLV2}}$ & $\phi_t  +\frac{1}{2}(\nabla\phi)^2 + g\eta = - 2\nu\frac{1}{\eta_x^2 + 1}\left(\phi_{zz}+W_z\right)- \frac{P_\textrm{rot}}{\rho} $  &  $\phi_t  + \frac{1}{2}(\nabla\phi)^2+ g\eta^*  = -4\nu \phi_{zz}$
\\
$z< \textrm{upper boundary}$  &   NS  &   &  $\nabla  P_{\text{rot}} = U_t - ((\vec{U}+\nabla \phi)\cdot \nabla)\vec{U)}-(\vec{U}\cdot \nabla\phi)\nabla\phi + \nu \nabla^2\vec{U}$ &  \\ \hline
\end{tabular}}\caption{Summary of equations for systems A,B and C.}\label{tab:summary} }
\end{sidewaystable}

\subsection{Nonlinear vortical terms}\label{sec_Disc_NLV}
Comparing the nonlinear vortical version of System A (Eqs (\ref{sys_A_total})) to the viscous water wave models presented in RFF and DDZ, the latter two ignore the nonlinear vortical terms: $\eta_x U$ in the KBC, and the $\phi_xU$ in the DBC. These terms are $\mathcal{O}(\eps \delta k)$. RFF justifies neglecting these terms because both the steepness, $\eps$, and the thickness of the boundary layer, $\delta k$, are very small quantities and thus their product leads to a negligible contribution.

However, a simple order analysis in \ref{app_OrderAnalysis} shows that the nonlinear vortical terms $\eta_x U$ in the KBC and $\phi_xU$ in the DBC are larger than the linear viscosity terms $W$ or $2\nu \phi_{zz}$ when $\frac{\eps}{\delta k}=\frac{a}{\delta}>1$, which holds in most physical cases (see Figure \ref{fig_MagnVort}), a fact that is also remarked in \citep{Longuet-Higgins1960}. In Figure~\ref{fig_LengthScales}, the case $\frac{a}{\delta}>1$ is indicated by the dashed line, and the case $\frac{a}{\delta}<1$ by the dotted line.

Yet, when we construct a viscous Dysthe equation through the method of multiple scales \textit{including} the nonlinear vortical terms, they cancel out  (see \ref{app_ExpressionVort}). Thus, up to $O(\eps^4)$ in the MMS, the approximate models of RFF and DDZ give the same result as the nonlinear vortical variant of System A expressed in potential terms.
\bigbreak

\subsection{Comparison of the three systems}
First, the domain of application is different for each model. Systems A and C are written in potential flow terms, and are therefore suitable to obtain a viscous propagation equation for the envelope, by means of for instance the Method of Multiple Scales (MMS). Taking the MMS expansion to $\mathcal{O}\left(\eps^4\right)$, both systems A and C reduce to the viscous Dysthe equation \citep{Carter2016}. In addition, the DDZ version of System A has been used to model the effect of the eddy viscosity in breaking waves \cite{Tian2010}, integrating it using the pseudo-spectral method \citep{Choi1995}, where the value of $\nu$ now represents the eddy viscosity instead of the kinematic viscosity. However, in these potential flow descriptions of System A and System C, the details of the boundary layer are lost. When the boundary layer is of interest, System A in the RFF description, or System B can be of use.

In contrast to Systems A and C, no approximation is made in System B (Dommermuth). However, System B cannot provide an envelope equation. Indeed, as its original purpose was the study of vortical bores, it is well suited for a domain where the vortical part of the velocity vector plays an important role in the whole domain or is the subject of interest. However, this is obtained at the cost of a higher numerical complexity. It has to be solved using a numerical finite difference scheme, combining Fourier techniques and LU decomposition.

Secondly, our comparison also illustrates the link between the vortical part of the velocity vector and the rotational pressure. To demonstrate the equivalence between Systems A and C, \citet{Longuet-Higgins1992} rewrites the \textit{linearized } System A into the \textit{linearized } System C, by using $\eta=\eta^*+\eta'$. This demonstrates that the vortical terms in the free surface boundary conditions of System A can indeed be interpreted as an additional pressure $P_\delta$. In \ref{app_LH_NLRuvinsky} we derive that the \textit{nonlinear} versions of Systems A and C are equal if terms of order $\mathcal{O}(\eps^2)$ and $\mathcal{O}(\eps \delta k)$ can be neglected.

Similarly, the terms labeled 'vorticity terms in the boundary layer' in the integral in  Eq.~(\ref{eqn_Ptot_Ruv}) in System A, are equal to $\nabla P_{\text{rot}}$ in Eq.~(\ref{sys_B_Prot}) in System B. This equivalence indicates that these vorticity terms (which also contain mixed terms with $\phi$) in the boundary layer can be interpreted as the effect of the rotational pressure. In System A, the integral is only over the boundary layer. The net effect of the terms is very small, and, as discussed, these terms can be neglected. The contribution of the rotational pressure is already taken into account by the vortical term in the KBC (Eq.~(\ref{sys_A_KBC})). In System B, this contribution \emph{is} significant, as the equation for the rotational pressure spans the entire vertical domain $z \in (-\infty, \eta)$. 

Comparing the DBC's of Systems B and C illustrates that vorticity and added pressure $P_\delta$ play the same role. However, instead of using the fully nonlinear equation for the vorticity as in System B (Eq.~(\ref{sys_B_Prot})), System~C \citep{Longuet-Higgins1969,Longuet-Higgins1992} relies on the physical argument of mass influx, using only \textit{linear} equations to obtain the additional pressure, as shown in Section \ref{sec_sys_LH}.

\section{Conservation of mass} \label{sec_ConsMass}
\begin{figure}[ht]
    \centering
\includegraphics[width=0.7\textwidth]{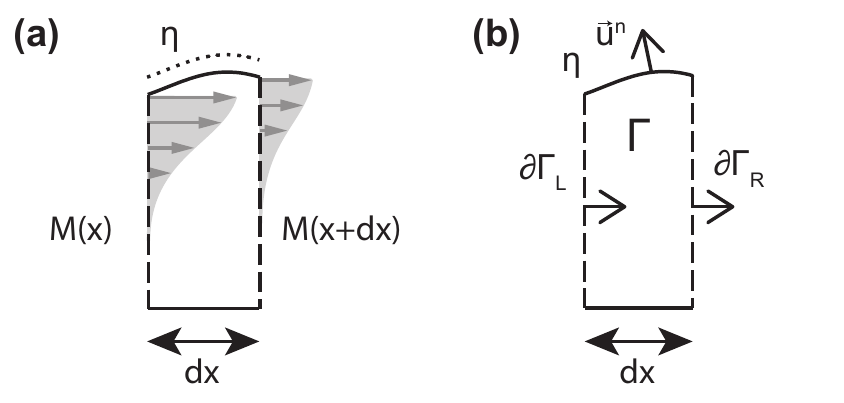}
\caption{(a) The difference in mass-flux $M$ (grey shaded area) will lead to a rise or fall of the surface elevation over the interval $dx$. (b) Domain of mass conservation} \label{fig_MassCons}
\end{figure}

Since the three systems deal differently with the distinction between the irrotational bulk and the rotational boundary layer, we verify the mass conservation condition for the entire domain. As we consider the fluid to be incompressible, this of course reduces to the conservation of volume. We examine the continuity equation ($\nabla \cdot\vec{u}=0$) on the domain $\Gamma$ in Fig.~\ref{fig_MassCons}b, with depth $z \in (-\infty, \eta)$ and width $dx$. 

The difference between the flux through side boundaries $\partial \Gamma_L$ and $\partial \Gamma_R$ denotes a total increase or decrease of fluid volume (area in 2D) in the domain in Fig.~\ref{fig_MassCons}a. In an incompressible fluid, this must be accommodated by an upward or downward movement of the free top boundary $\eta$. That is, the mass-flux through the side boundaries is compensated by the movement of the surface elevation $\eta$. 

First, the continuity equation is integrated over the column  $z \in (-\infty, \eta)$
\begin{align} \label{eqn_MassConsStep1}
\int_{-\infty}^{\eta(x)} (u_{x} + w_{z}) dz & = 0 .  
\end{align}
\noindent Using the Leibniz rule gives
\be \label{eqn_MassConsStep2}
\underbrace{\frac{d}{dx} \bigg(\int_{-\infty}^{\eta(x)} u dz}_\textrm{net in/out flux}\bigg)  \underbrace{-u(\eta)\frac{\partial \eta}{\partial x} + w(\eta)}_\textrm{$\tilde{u}^n @ \eta$} = 0 . 
\ee

\noindent Here, $\int_{-\infty}^{\eta(x)} u dz$ denotes the flux through a vertical boundary at a given position $x$, see Figure \ref{fig_MassCons}a. The sign of its derivative $d/d x$ from one position to the next shows volume coming into, or leaving the slice. The rate at which this increase in volume occurs is equal to the velocity of the boundary moving up or down to accommodate the change: $\vec{u}^n$. Here $\tilde{u}^n$ refers to $\vec{u}^n$ in Cartesian coordinates:  $\tilde{u}^n = \sqrt{\eta_x^2 +1}\vec{u}^n$ (see Eq.~\ref{eqn_KBC_u2}). Using the KBC (Eq.~(\ref{eqn_KBC_u})), we can write

\be \label{eqn_massConsv_v1}
\underbrace{\frac{d}{dx} \bigg(\int_{-\infty}^{\eta(x)} u dz\bigg)}_\textrm{Total flux} + \underbrace{\eta_t}_\textrm{Change of $\eta$ in time}  = 0 . 
\ee

\noindent This notation in terms of the velocity vector is a level of description higher than the one used in the three systems described above, {\it i.e.} before the introduction of the potential framework or the gauge choice of $\vec{U}$. 

\subsection{Comparison to the irrotational system}\label{sec_MC_irr}
First, we would like to address the use of the irrotational fluid as a benchmark for the conservation of mass of a viscous system. For the irrotational water wave problem, the potential notation  $\vec{u}=\nabla \phi_\textrm{\text{irr}}$ can be used in  Eq.~(\ref{eqn_massConsv_v1}), to obtain:

\be \label{eqn_massConsv_irr}
\underbrace{ \frac{d}{dx} \bigg(\int_{-\infty}^{\eta(x)}  \phi_{\textrm{\text{irr}},x}  dz\bigg) }_\textrm{Total flux} + \underbrace{ \eta_{t}}_\textrm{Change of $\eta$ in time}   = 0 . 
\ee

For the viscous water wave problem, when using the Helmholtz decomposition for $\vec{u}$ in  Eq.~(\ref{eqn_massConsv_v1}), this yields 

\be \label{eqn_massConsv_rot1}
\frac{d}{dx} \bigg(\int_{-\infty}^{\eta(x)}(\phi_{x} +U) dz\bigg) + \eta_t = 0. 
\ee

Rewriting $\int U dz$ using the Leibniz rule and the continuity equation ($U_x = - W_z$), gives

\begin{align}\label{eqn_massConsv_rot}
\begin{split}
  \underbrace{\frac{d}{dx} \bigg(\int_{-\infty}^{\eta(x)} (\phi_{x}) dz\bigg)  }_{\textrm{Irrotational flux}}  - \underbrace{(-U\eta_x + W)}_\textrm{${U}^n$}|_{z=\eta}  
  &+ \underbrace{\eta_t }_{\textrm{Change of $\eta$ in time}} =0,
\end{split}
\end{align}
\noindent where ${U}^n$ is the normal component of the vortical velocity. Using for explicit expressions for the vortical terms, for instance from the DDZ system (Eq.(\ref{sys_DDZ_KBC})), where $-U\eta_x = 0$ and $W=2\nu \eta_{xx}$, results in
\begin{align}\label{eqn_massConsv_Dias}
\begin{split}
\frac{d}{dx}\bigg(\int_{-\infty}^{\eta(x)} (\phi_{\textrm{DDZ},x}) dz\bigg)   \\  \underbrace{-2\nu\eta_{xx}}_\textrm{extra term}
  &+ \int_{x_0}^{x_0 + \Delta x} \eta_t   dx  =0 .
\end{split}
\end{align}

In comparison to Eq. (\ref{eqn_massConsv_irr}) the additional term $2\nu\eta_{xx}$ seems to break the conservation of mass. However, as pointed out in Sec.~\ref{sec:ShearStress}, it is unphysical to have a viscous irrotational fluid with a curved free-surface boundary: a viscous fluid \textit{cannot} be irrotational at the boundary. Therefore, one must remember that $\eta_t$  is based on the rotational KBC (Eq.(\ref{sys_DDZ_KBC})):

\begin{align}\label{eqn_massConsv_Dias2}
\begin{split}
  \underbrace{ \frac{d}{dx} \bigg(\int_{-\infty}^{\eta(x)} (\phi_{\textrm{DDZ},x}) dz\bigg)}_{ \textrm{Irrotational flux}}  \\ \underbrace{\cancel{-2\nu\eta_{xx}}}_\textrm{extra term}  
  &+ \left(\underbrace{ -\eta_x\phi_x|_{z=\eta}  + \phi_z|_{z=\eta}  }_{\textrm{irrotational } \tilde{\eta}_t} + \cancel{2\nu \eta_{xx}} \right)   =0 .
\end{split}
\end{align}

The irrotational mass flux due to $\phi_{\textrm{DDZ}}$ is equal to the irrotational $\tilde{\eta_t}$. This illustrates that the rotational and irrotational parts of the problem superpose linearly, as in the formulation of the Helmholtz equation. We therefore confirm that the mass is conserved in the DDZ system\footnote{Examining the integrability of the DDZ system, \cite{Ngom2018} points out that when the system is assumed periodic, the 'extra term' disappears. This is indeed valid for their system of periodic functions and their following development. However, here,  we want to consider the conservation of mass on any domain, not just periodic ones.}.

\subsection{Mass conservation in the three systems}
In the following we compare the expressions for conservation of mass for systems A, B and C. We write the condition for mass conservation such that $\nabla\phi$ is the sole contributor to the mass-flux on the side boundaries, and any vorticity terms are expressed on the moving upper boundary.

\subsubsection{System A}\label{sec_MC_A}
The Helmholtz decomposition is used for $\vec{u}$ in  Eq.~(\ref{eqn_massConsv_v1}), and results in Eq.~(\ref{eqn_massConsv_rot}), repeated below for clarity:
\be \label{eqn_massConsv_A}
\underbrace{\frac{d}{dx} \bigg(\int_{-\infty}^{\eta(x)}  \phi_{\textrm{A},x} + U_{\textrm{A}} dz\bigg) }_\textrm{Total flux} + \underbrace{ \eta_t }_\textrm{Change of $\eta$ in time }   = 0 . 
\ee

\subsubsection{System B}
We repeat the same exercise for System B, however Eq.~(\ref{sys_B_KBC1}) is irrotational, indicated with the tilde. The mass conservation Eq.~(\ref{eqn_massConsv_rot1}) becomes  
\be \label{eqn_massConsv_B}
\underbrace{\frac{d}{dx} \bigg(\int_{-\infty}^{\eta(x)}  \phi_{\textrm{B},x} + U_{\textrm{B}} dz\bigg) }_\textrm{Irrotational flux} + \underbrace{ \tilde{\eta}_t }_\textrm{irrotational}   = 0 . 
\ee

\subsubsection{System C}
Again, the same method is repeated for System C. The conservation of mass, Eq.~(\ref{eqn_massConsv_v1}), is now written for both domains; the irrotational bulk $z \in (-\infty, \eta^*)$ ($\Gamma_I$), and the rotational boundary layer $z \in (\eta^*, \eta)$ ($\Gamma_{II}$):

\be \label{eqn_massConsv_LH1}
   \underbrace{ \frac{d}{dx} \bigg(\int_{-\infty}^{\eta^*(x)} \phi_{\textrm{C},x}  dz\bigg)  +  \eta^*_t}_\textrm{Domain $\Gamma_{I}$} +    \underbrace{  \frac{d}{dx} \bigg(\int_{\eta^*(x)}^{\eta(x)} U_\text{C}  dz\bigg)  +  \eta'_t }_\textrm{Domain $\Gamma_{II}$} =0 .
\ee
Recall from Section \ref{sec_sys_LH} that in $\Gamma_{II}$, by definition, the velocity vector has only the rotational component $\vec{U}$. Therefore $\phi_C=0$ in $\Gamma_{II}$. Similarly, $U_{\textrm{C}}=0$ in $\Gamma_{I}$. Consequently, both can be taken together in one integral that spans both domains, i.e. $z \in (-\infty, \eta)$. In addition, we rewrite $\eta= \eta^*+\eta'$ to obtain
\be \label{eqn_massConsv_LH}
\underbrace{\frac{d}{dx} \bigg(\int_{-\infty}^{\eta(x)}  \phi_{\textrm{C},x} + U_{\textrm{C}} dz\bigg) }_\textrm{Total net-flux} + \underbrace{\eta_t}_{\textrm{rotational}}  = 0 . 
\ee

\subsection{Mass conservation comparison}\label{sec_MassConsvDisc}

We posed the conservation of mass condition for each system using the Helmholtz notation. As such, the conservation of mass can be posed individually for the rotational and irrotational part of the velocity vector. Using the explicit definition of the vortical terms in the DDZ version of System A, we show in Eq.~(\ref{eqn_massConsv_Dias2}) that this system conserves mass. By the same argument as in Section \ref{sec_MC_irr}, the soundness of the nonlinear vortical version and the RFF version of System A can be established. System C is by definition  (Eq.~(\ref{eqn_massConsv_LH1})) an explicit addition of the irrotational bulk and the rotational boundary layer. For System B (Eq.~(\ref{eqn_massConsv_B})), like for the other two systems, the mass conservation condition in will be fulfilled provided no numerical errors are introduced in obtaining concrete values for the rotational velocity vector $\vec{U}$ and velocity potential $\phi$.

Furthermore, comparing the mass conservation conditions for Systems A (Eq.~(\ref{eqn_massConsv_A})) and C (Eq.~(\ref{eqn_massConsv_LH})), we can deduce that $\nabla\phi_C=\nabla\phi_{\text{irr}}=\nabla\phi_A$\footnote{Note that Eq.~(\ref{eqn_massConsv_LH}) corresponds to the fully nonlinear version of the model, i.e. before any form of linearization is introduced, since the normal vector $\vec{U}^n$ is not yet expressed in terms the vortical pressure $P_\delta$. }. Therefore, using a purely irrotational system for the viscous water wave problem neglects the vortical contribution to the mass-flux represented by the term $\int_{\eta^*(x)}^{\eta(x)} U  dz$ in Eq.~(\ref{eqn_massConsv_LH1}), which is the definition of the vortical layer in Eq.~(\ref{eqn_LH_Un}). This is the origin of the pressure term $P_{\delta}$ in System C (Section \ref{sec_sys_LH}). When this additional pressure term is missing in the model equations, it leads to an underestimation of the decay rate, as exemplified in \citet{Padrino2007}.

Finally, for System B (Eq.~(\ref{eqn_massConsv_B})) both the rotational and irrotational part of the horizontal velocity $u$ are needed to account for the movement of the \textit{irrotational} $\tilde{\eta}$ (irrotational KBC). In contrast,  
System A (Eq.~(\ref{eqn_massConsv_A})) and System C (Eq.~(\ref{eqn_massConsv_LH})), concern the movement of $\eta$ based on the rotational KBC. This exemplifies the gauge choice for System B where the vortical part of the velocity $\vec{U}$ also takes part of the irrotational velocity in the bulk, as indicated by the grey shaded area in Figure \ref{fig_3situations}B.

\section{Discussion and Conclusion}
In summary, we contrast and compare three different ways of looking at the viscous water wave problem, both in the model equations and in the resulting conservation of mass.  

Practical implementation is the guiding factor in opting for one or the other. System B (\citet{Dommermuth1993}) is the only closed exact model that is fully nonlinear without approximations, at a cost of having more complicated  equations in the bulk, where the fluid is assumed to be rotational, as well as an additional equation for the rotational pressure. To close the systems A and C, linearization has to be applied, and approximations have to be made.  System C and the DDZ version of System A do not provide details on the boundary layer but are computationally simple, and can provide a basis for envelope evolution equations like the Nonlinear Schr\"odinger or Dysthe equations. \linebreak

We verify for all three systems that the mass conservation conditions are in agreement with the Helmholtz superposition of the rotational and irrotational flow. The comparison of the mass conservation conditions illustrates that the gauge choice of System B requires the vortical velocity vector to contain also an irrotational part.

It would be interesting to see how well the linear simplification of the vortical pressure captures the dynamics for the total pressure, comparing to results based on the fully nonlinear model by \citet{Dommermuth1993}. 
Various other models hint at the existence and relevance of nonlinear viscosity terms in the propagation equation \citep{Armaroli2018_visc,Fabrikant1980}. The order analysis in Cartesian coordinates performed in \ref{app_OrderAnalysis} and a leading order analysis in curvilinear coordinates performed by \citet{Phillips1979} show their relevance in different regimes of viscosity and wave steepness. However, we show that, when the system is closed in potential terms so that the multiple-scale method can be applied, such viscosity terms cancel out each other and have no effect on the viscous Dysthe equation.  

Depending on the physical application, it is important to have a clear understanding of the limitations and strength of each possible formulation of the viscous free-surface problem. We hope that this comparison contributes to a clearer interpretation of this problem, and will be built upon in future analyses.

\section*{Acknowledgements}
We acknowledge the financial support from the Swiss National Science Foundation (Projects Nos. 200021-155970 and 200020-175697). We would like to thank John Carter, Peter Wittwer and Yves-Marie Ducimeti\`ere for fruitful discussions.

\bibliographystyle{model1-num-names}
\bibliography{Vorticity.bib}


\newpage
\appendix
\newpage


\section{Nonlinear vortical terms}
\subsection{Closing system A}\label{app_ExpressionVort}
In order to close System A, the rotational part of the velocity, $\vec{U}$, needs to be expressed in terms of either $\phi$ or $\eta$. By solving for the linearized Navier-Stokes (NS) equations, \citet{Dias2008} (DDZ) show $W=A_x~=~2\nu\eta_{xx}$ in the linear viscous water wave problem, and conjectures that this also holds for the nonlinear case. We can follow the same line of reasoning to find an expression for $U=A_z$. 
Following \citet{Lamb1932,Wang2006} and DDZ, using the linearized NS, 
the solution where both $\phi$ and $A$ are periodic in $x$ and must have the form
\be
\begin{split}\label{def_PhiA}
\phi (x,z,t) &= \phi_0 e^{i(kx-\omega t)}e^{|k|z} \\
A (x,z,t) &= A_0 e^{i(kx-\omega t)}e^{mz} 
\end{split}
\ee
\noindent 
yielding the relation
\be
\begin{split}\label{eqn_m}
m^2 &= k^2-i\frac{\omega}{\nu}.
\end{split}
\ee
If the nonlinearity of the waves plays an important role, the expression obtained for the potential vector $A$ in the \textit{linear} equations is not sufficient. Instead, the nonlinear system should be used. To simplify the exponential behavior in $z$, we can consider that from Eq.~(\ref{eqn_m})

\be
\begin{split}\label{eqn_mparts1}
\Re(m) &= +\frac{1}{\sqrt{2}}\sqrt{\sqrt{k^4+\frac{\omega^2}{\nu^2}}+k^2}\\
\Im(m) &= -\frac{1}{\sqrt{2}}\sqrt{\sqrt{k^4+\frac{\omega^2}{\nu^2}}-k^2}
\end{split}
\ee

\noindent and realizing that for typical values of $k$ the relation $\frac{\omega}{\nu} = \frac{2}{\delta^2} \gg k^2$ holds, so that Eq.~(\ref{eqn_mparts1}) reduces to

\be
\begin{split}\label{eqn_mparts2}
\Re(m) &\approx +\frac{1}{\delta}\\
\Im(m) &\approx -\frac{1}{\delta}
\end{split}
\ee

To derive an expression for $A_z$ from Eq.~(\ref{def_PhiA}), we consider only the real part of $m$ (Eq.~(\ref{eqn_mparts2})), since it is part of a real valued velocity vector \citep{Lamb1932}, and obtain
\be
A_z\Big|_{z=0}  = 2\sqrt{\frac{\nu}{2\omega}}\omega\eta_x= -2 \sqrt{\frac{\nu}{2\omega}}\phi_{zz} =  -\frac{2k}{\sqrt{2}}  \sqrt{\frac{\nu}{\omega}}\phi_{z}
\ee

Note that like DDZ, we \textit{conjecture} that the linear versions of $\phi$ and $A$ are sufficiently precise to remain valid in the nonlinear KBC and DBC.  The closed form of System A, including the nonlinear vortical terms can be written as:

\begin{subnumcases}{\label{sys_EulerVortPhi}\hspace{-3em} } 
\phi_{xx} + \phi_{zz} = 0 &\hspace{-2em}  $-\infty<z<\eta$   \\
\nabla \phi \to 0 & \hspace{-2em} $z\to-\infty$ \\
\eta_t + \phi_x \eta_x -\underbrace{2\sqrt{\frac{\nu}{2\omega}}\phi_{zz}\eta_x}_{\text{NLV1}}   =\\ \nonumber
\hspace{8em} \phi_z + {2\nu \eta_{xx}}    &\hspace{-2em} KBC  \label{A_sys_A_NS}  \\
\phi_t + \frac{1}{2}\left(\phi_x^2 + \phi_z^2 \right) + g\eta   = \\ \nonumber
\hspace{4em} - 2 \nu \phi_{zz}  + \underbrace{2 \sqrt{\frac{\nu}{2\omega}}\phi_{zz}\phi_x}_{\text{NLV2}}  &\hspace{-2em} DBC \label{A_sys_A_DBC} 
\end{subnumcases}

where the nonlinear vortical terms (NLV1, NLV2) correspond to those in System A~(Eqs.~(\ref{sys_A_total})).

\subsection{Propagation equation}
Using the method of multiple scales (MMS), as described in \citet{Carter2016}, gives  the following propagation equation for the envelope $B$:

\begin{multline}\label{eqn_propeqn_vort}
\frac{\partial B}{\partial t} + \frac{\omega_0}{2k_0} \frac{\partial B}{\partial x} = 
 \eps\bigg[ +i\frac{\omega_0}{8 k_0^2}  \frac{\partial^2 B}{\partial x^2} + \frac{1}{2}i k_0^2 \omega_0 B|B|^2 -2 k_0^2\nu B \bigg]  \\
 + \eps^2\bigg[ - \frac{3}{2} k_0 \omega_0 |B|^2\frac{\partial B}{\partial x} -  \frac{1}{4}k_0 \omega_0 B^2 \frac{\partial B^*}{\partial x}  \\ 
+ \frac{\omega_0}{16k_0^3}\frac{\partial^3 B}{\partial x^3} +i k_0 B\frac{\partial \phi}{\partial x} - 4i k_0 \nu\frac{\partial B}{\partial x}   \\
\underbrace{-2 k_0^3\omega_0 \delta B|B|^2 }_{\text{NLV1}} + \underbrace{2 k_0^3\omega_0 \delta B|B|^2 }_{\text{NLV2}}\bigg]
\end{multline}

The nonlinear vorticity terms cancel out. Hence, the solution reduces to the one given by the same system proposed by DDZ. \citet{Zhang1997} indeed also remark that RFF only take into account the linearized vorticity normal vector, and that the linear vorticity terms describe experiments in reasonable agreement.

\subsection{Order analysis} \label{app_OrderAnalysis}
To check for the relevance of the nonlinear vortical terms in the KBC and DBC the following order analysis is performed. Following Eq.~(\ref{eqn_HH}), we write $U=-A_z$ and $W=A_x$. Note that from the expression of $A$ in \ref{def_PhiA}, $A \propto e^{\delta^{-1}z}$, and thus decays from a finite value to a negligible value over a vertical distance $\delta$. However, $\nabla\phi$ is not affected by viscosity and experiences an exponential decay over a characteristic length $k^{-1}$.

Also note that the analysis in Cartesian coordinates is only valid if $\eps \ll 1$. The variables are scaled as follows

\begin{align}\label{sys_Scaling}
x&=k^{-1}  \tilde{x}         &  \phi &=  \phi_0 \tilde{\phi}  \nonumber \\
t&=\omega^{-1}\tilde{t}      &  A &= A_0  \tilde{A}  \nonumber    \\
z&=k^{-1}\tilde{z}  \qquad \textrm{for } \phi       &  \eta &= a \tilde{\eta}\\
z&=\delta\tilde{z}  \qquad  \quad  \textrm{for } A   &  \nonumber
\end{align}

\noindent where $a,A_0$ and $\phi_0 =  \frac{a \omega}{k}$ are the initial amplitudes of the corresponding quantities. Note that,  following \citep{Dias2008}, if the viscosity is small $\Theta = \frac{A_0}{\phi_0} \approx \frac{\nu k^2}{\omega} \ll 1$.  The KBC (Eq.~(\ref{sys_A_KBC})) and  DBC (Eq. ~(\ref{sys_A_DBC})) can be written as

\begin{figure} [h]
\centering
\includegraphics[width=0.45\textwidth]{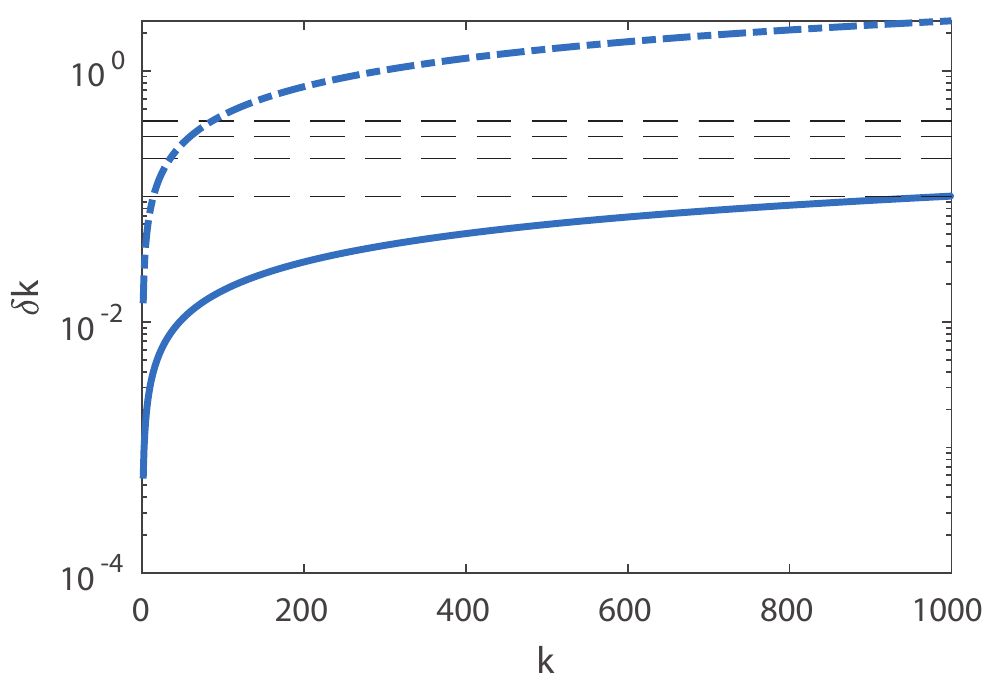}
\caption{$\delta k$ as a function of wave number $k$ for the kinematic viscosity of water, $\nu= 10^{-6}\mathrm{\,m^2\,s^{-1}}$, (solid line) and for glycerine, $\nu=6.21\times 10^{-4}\mathrm{\,m^2\,s^{-1}}$,  (dashed-dotted line). The steepness values $\eps$=0.1, 0.2, 0.3, 0.4 are indicated with horizontal dashed lines. Above $\eps$=0.44 waves are generally considered to break \citet{Toffoli2010a}.}\label{fig_MagnVort}
\end{figure}

\begin{equation} \label{eqn_KBVDBC_adim}
\begin{aligned}
 \tilde{\eta}_{\tilde{t}} + \eps \tilde{\phi}_{\tilde{x}} \tilde{\eta}_{\tilde{x}} -\tilde{\phi}_{\tilde{z}}   &=  (\delta k)^2\tilde{A}_{\tilde{x}} + \eps\delta k\tilde{A}_{\tilde{z}}\tilde{\eta}_{\tilde{x}} & & \textrm{KBC}	\\
 \tilde{\phi}_{\tilde{t}} + \eps\frac{1}{2}(\tilde{\nabla}\tilde{\phi})^2 - \tilde{\eta} &= -2 (\delta k)^2 \tilde{\phi}_{\tilde{z}\tilde{z}} -    \eps \delta k \tilde{\phi}_{\tilde{x}} \tilde{A}_z & & \textrm{DBC}
\end{aligned}
\end{equation}

\noindent using $U=-A_z$ and $W=A_x$, it is clear that if
\be\label{eqn_MangVortTerms}
\eps > \delta k \quad  \Rightarrow \quad  \eta_x U  > W, \quad  \phi_x U > \phi_{zz}
\ee

Therefore, at fixed $k$, the nonlinear vorticity terms become more relevant for steeper waves or weaker viscosity. Fig.~\ref{fig_MagnVort} shows the value of $\delta k$ for water (solid line) and glycerine (dashed line), as a function of wave number. Comparing this to the wave steepness, indicated by the horizontal dashed lines, shows that for practically all wave-numbers $ \eps > \delta k$ in water. The relative magnitudes found in our dimensional analysis (Eq.~(\ref{eqn_MangVortTerms})) are  is confirmed by the leading order expressions for the rotational and irrotational velocity vector in curvilinear coordinates (\citep{Phillips1979}, Chapter 3). \citet{Longuet-Higgins1953} indeed also mentions that $\eps/  \delta k  = a/ \delta \gg 1$ is the common physical situation.

For this reason, we retained the nonlinear vortical terms in Eqs. (\ref{sys_A_total}), labeled VNL1 and VNL2. Surprisingly however, these terms compensate each other in the evolution equation for the envelope, Eq.~(\ref{eqn_propeqn_vort}). In particular, they give a null contribution at the 4$^{th}$ order level in steepness  \footnote{Performing the order analysis on Eqs.~(\ref{sys_EulerVortPhi}), using the same scaling as in Eq.~(\ref{sys_Scaling}), indeed gives the same adimensional equation as Eq.~(\ref{eqn_KBVDBC_adim})}.

 \section{Conversion of System A to System C} \label{app_LH_NLRuvinsky} 
\citet{Longuet-Higgins1992} demonstrates that the linearized system presented in RFF
\begin{equation}
\begin{aligned}
\eta_t   & =  \phi_z + W  	\\
\phi_t  + g\eta  & =   -  2 \nu \phi_{zz}
\end{aligned}
\end{equation}

\noindent can be written, using $\eta = \eta^* + \eta'$, as

\begin{equation}\label{eqn_app_LH_A_lin}
\begin{aligned}
\eta_t  + \eta^*_x\phi_x  + \eta'_x\phi_x &=  \phi_z \\
\phi_t  + \frac{1}{2}(\nabla\phi)^2+ g\eta^*  & =   -  4 \nu \phi_{zz}
\end{aligned}
\end{equation}

\noindent where the boundary layer induced-pressure method described in \citet{Longuet-Higgins1969} is used. We perform the same exercise for the nonlinear System A. 

\subsubsection*{The kinematic boundary condition}

Since  $\eta'_t = -\eta_x U + W$, Eq.~(\ref{sys_A_KBC}) can be rewritten as
\begin{equation}
\begin{aligned}
\eta^*_t +  \eta^*_x\phi_x + \eta'_x\phi_x &=  \phi_z   
\end{aligned}
\end{equation}

\noindent In \citet{Longuet-Higgins1969}, the nonlinear term $\eta'_x\phi_x$ is ignored. Since $\eta^*_x\propto a$ and $\eta'_x \propto \delta$, this is justified if $a \gg \delta$ (or $\eps \gg\delta k$), which looking at Figure \ref{fig_MagnVort} is the case in most physical situations. It can also be solved by moving to tangential coordinates, and recalling Eq.~(\ref{eqn_KBC_u3}), and writing
\begin{equation}
\begin{aligned}
\eta^*(s,t)_t + \eta'(s,t)_t &= u^n(s,t) 	\\
\eta^*_t &=  \phi_n\\
\end{aligned}
\end{equation}
\noindent 
resulting indeed in the irrotational KBC as in System C. 

\subsubsection*{The dynamic boundary condition}
Starting from Eq.~(\ref{eqn_Ptot_Ruv}), we insert the normal stress balance, and replace $\eta = \eta^* + \eta'$, to obtain

\begin{multline} \label{A_eqn_Ruvinsky_DBC1}
\phi_t + \frac{1}{2} (\nabla \phi)^2 +   2 \nu \left(\phi_{nn} + U^n_n \right)  - g\eta^*  -g\eta' = \\\underbrace{ - \int_{\eta-\delta}^{\eta} \left( \vec{U}_t- \nu \nabla^2 \vec{U}   - \omega \times \vec{U} \right) dz   - \frac{1}{2}\vec{U}^2}_\textrm{Vortical layer} \\
\underbrace{\phi_n U^n + \phi_s U^s}_\textrm{Mixed NL terms} \qquad 
\end{multline}

Recall the definition of the vortical layer in system C:
\begin{equation}\label{eqn_etaprime}
\begin{aligned}
\eta' &= \int U^n dt\\
  &= -\int \eta_x U dt+\int W dt\\
  &= -\int \eta_x U dt- 2\nu \frac{k}{\omega^2} \phi_{zz}
\end{aligned}
\end{equation}

This expression agrees with the one obtained in \citet{Longuet-Higgins1992} in the linear limit (see Eq. (3.11) in that paper). Inserting Eq.~(\ref{eqn_etaprime}) into Eq.~(\ref{A_eqn_Ruvinsky_DBC1}) and denoting VL for the small terms in the vortical layer:

\begin{multline} \label{A_eqn_Ruvinsky_DBC2}
\phi_t + \frac{1}{2} (\nabla \phi)^2 +   2 \nu \left(\phi_{nn} + U^n_n \right)  - g\eta^*   = \\
-  g\int \eta_x U dt - 2\nu \phi_{zz} + \text{VL} + \phi_n U^n + \phi_s U^s \qquad 
\end{multline}

Taking $z\approx n$ for $\phi$, which is justified for $\eps\ll1$ gives
\begin{multline}  \label{eqn_app_LH_DBC3}
\phi_t  + \frac{1}{2}(\nabla\phi)^2-g\eta^*   = - 4\nu \phi_{zz} \\
- 2\nu\vec{U}^n_n- g\int \eta_x U dt + VL + \phi_n U^n + \phi_s U^s 
\end{multline}
Systems A and C are equal if terms of $\mathcal{O}(\eps^2)$ and $\mathcal{O}(\eps \delta k)$, {\it i.e.} the second line of the equation above, can be ignored.

\end{document}